\documentclass[twocolumn,prl,showpacs,preprintnumbers,amsmath,amssymb]{revtex4}
\usepackage{graphicx,epsfig}
\usepackage{dcolumn}
\usepackage{bm}
\begin{document}
\title{Anomalous glassy relaxation near the isotropic-nematic phase 
transition}
\author{Prasanth P. Jose, Dwaipayan Chakrabarti and Biman Bagchi} 
\email{bbagchi@sscu.iisc.ernet.in}
\affiliation{Solid State and Structural Chemistry Unit, Indian Institute of 
Science, Bangalore 560012, India}

\date \today

\begin{abstract}

{\it Dynamical heterogeneity} in a system of Gay-Berne ellipsoids near 
its isotropic-nematic (I-N) transition and also in an equimolar mixture of 
Lennard-Jones spheres and Gay-Berne ellipsoids in deeply supercooled regime
is probed by the time evolution of non-Gaussian parameter (NGP). The 
appearance of a dominant second peak in the rotational NGP near the 
I-N transition signals the growth of pseudonematic domains. Surprisingly 
such a second peak is instead observed in the translational NGP for glassy 
binary mixture. Localization of orientational motion near the I-N transition
is found to be responsible for the observed anomalous orientational 
relaxation.
 
 
\end{abstract}

\maketitle

%\large

The dynamics of liquid crystals \cite{deGennes-book} across the 
isotropic-nematic (I-N) phase transition displays an array of dynamical 
features that have been a subject of renewed interest in recent times 
\cite{Cang-Li-Novikov-Fayer-JCP-2003, Gottke-Brace-JCP-2002, 
Gottke-Cang-JCP-2002, Jose-Bagchi}. In particular, the 
emergence of multiple time scales in orientational relaxation has been 
demonstrated in the isotropic phase of a number of liquid crystals as the 
I-N transition temperature $T_{IN}$ is approached from above 
\cite{Gottke-Brace-JCP-2002,Gottke-Cang-JCP-2002}. The long time exponential
decay with a temperature dependent time constant has been observed to obey 
the Landau-de Gennes (LdG) theory \cite{deGennes-book}. This long 
time decay is ascribed to the orientational randomization of the so 
called pseudonematic domains. Although the isotropic phase is 
macroscopically homogeneous, near the I-N transition a local nematic-like 
order persists over a length scale $\xi (T)$ that characterizes these 
pseudonematic domains. The LdG theory predicts the {\it correlation length} 
$\xi (T)$ to grow as temperature $T$ approaches $T_{IN}$ from above: 
$\xi (T) = \xi_{0} \left[T^{\star}/(T -T^{\star})\right]^{1/2}$ where $\xi_{0}$ is a molecular length. It suggest that $\xi (T)$ eventually diverges at 
$T^{\star}$ which falls just below $T_{IN}$ (typically, 
$T_{IN} - T^{\star} \sim 1~K$) \cite{deGennes-book}. The concomitant slow 
down of orientational relaxation is so dramatic that it finds an 
{\it analog} in the phenomenal increase of the time scale of orientational 
dynamics in the molecular liquids with the glass transition temperature 
approached from above \cite{Cang-Li-Novikov-Fayer-JCP-2003}.
  
Recently Fayer and coworkers have carried out extensive optical Kerr effect 
(OKE) measurements which indicate that the analogy between the isotropic 
phase of liquid crystals and the supercooled liquids in orientational 
relaxation {\it persists over a rather wide range of time scales}
\cite{Gottke-Brace-JCP-2002,Cang-Novikov-Fayer-JCP-2003,
Cang-Li-Novikov-Fayer-JCP-2003}. A description, that involves a power law at 
short times and an exponential decay at long times with a second power law 
in the crossover region, has been found to conform well to the optical 
heterodyne detected OKE experimental data for a number of liquid crystals
in the isotropic phase and supercooled liquids
\cite{Cang-Li-Novikov-Fayer-JCP-2003}. The resemblance suggests that
a similar underlying microscopic mechanism for relaxation is operative 
in liquid crystals in the isotropic phase near $T_{IN}$ and in the 
supercooled liquids even though the latter do not undergo a 
structural phase transition \cite{Cang-Li-Novikov-Fayer-JCP-2003}. 

The goal of the present Letter is to explore the analogy following a 
different approach. The appearance of pseudonematic domains in the isotropic 
phase near $T_{IN}$ is expected to result in heterogeneous dynamics. Such
heterogeneity in dynamical behavior has been observed for glassy liquids in 
a large number of experimental and computer simulation 
studies \cite{Ediger-ARPC, Sillescu-JNCS, Richert-JPCM}, but, to the best of 
our knowledge, has not been studied near the I-N transition. We have 
therefore carried out extensive computer simulation 
studies to gain further insight into the apparent analogy between the 
dynamics of the isotropic phase of the liquid crystals near $T_{IN}$ and 
that of the supercooled liquids from the perspective of heterogeneity with 
special emphasis on rotational degrees of freedom. To this end, we 
have investigated two systems: $(1)$ a collection of $576$ Gay-Berne 
ellipsoids (with aspect ratio $\kappa = 3$) along an isotherm at a series of 
densities across the I-N  transition; $(2)$ a binary mixture of $128$ 
Lennard-Jones spheres and $128$ Gay-Berne ellipsoids along an isochore at a 
series of temperatures down to the deeply supercooled regime. For our liquid
crystal system, we have used the well-known Gay-Berne pair potential 
\cite{Gay-Berne-JCP-1981-Brown-PRE-1998} with the most studied 
parametrization \cite{GB-parameter}, that is known to undergo a rather 
sharp I-N transition on scanning the density along an 
isotherm \cite{Miguel-Vega-JCP-2002}. In a 
recent simulation study \cite{Jose-Bagchi}, this system has been shown to 
exhibit a short time power law decay of the second rank orientational 
correlation function in single-particle as well as collective dynamics on 
approaching the I-N transition from the isotropic side. On the other hand, 
the binary mixture studied here is a {\it new} model introduced with an 
intention to study orientational dynamics in the supercooled regime. The 
choice of a binary system is motivated by the success of the Kob-Andersen 
binary mixture \cite{Kob-Andersen-PRE}, that is widely used for computer 
simulations of supercooled liquids as the additional complexity of the 
system is sufficient to avoid crystallization. Here, the interaction 
potential between a sphere and an ellipsoid, which is chosen to be a prolate (with $\kappa = 2$), is given by following Cleaver and 
coworkers \cite{Cleaver-PRE-JCP}. We have determined the set of energy and 
length parameters such that neither any phase separation occurs nor any 
liquid crystalline phase with orientational order appears even at the lowest temperature studied at a high density \cite{binary-parameter}. For the 
present purpose, we concentrate on the dynamics of only ellipsoids in the 
binary mixture.

This Letter invokes non-Gaussian parameters \cite{Rahman} for the 
description of dynamical heterogeneity \cite{Kob-Andersen-PRE,ngp}. The 
non-Gaussian parameter (NGP) for the translational degrees of freedom is 
defined for a three-dimensional system of $N$ rigid bodies of the same kind 
by
\begin{equation}
\alpha_{2}^{(T)} = \frac{3~<\Delta {\bf r}(t)^{4}>}
{5~<\Delta {\bf r}(t)^{2}>^{2}} - 1
\end{equation}
where $<\Delta {\bf r}(t)^{2n}> = \frac{1}{N}\displaystyle \sum_{i=1}^{N}
<|{\bf r}_{i}(t) - {\bf r}_{i}(0)|^{2n}>$. 
Here ${\bf r}_{i}(t)$ is the position of the center mass of the {\it i}th 
body at time $t$. If ${\bf r}(t)$ is a Gaussian process, $\alpha_{2}^{(T)}$ 
vanishes for all time. The analog to this description for the rotational 
degrees of freedom can be obtained through the replacement of
${\bf r}_{i}(t)$ by the corresponding angular variable ${\bm \phi}_{i}$ 
\cite{rd-PRE}, the change of which is defined in terms of angular
velocity ${\bm \omega}_{i}$ by $\Delta {\bm \phi}_{i}(t) = {\bm \phi}_{i}(t) - {\bm \phi}_{i}(0) = \int_{0}^{t} dt^{\prime} 
{\bm \omega}_{i}(t^{\prime})$. For the symmetry of the ellipsoids, the 
definition for the rotational non-Gaussian parameter then reads as
\begin{equation}
\alpha_{2}^{(R)} = \frac{<\Delta {\bm \phi}(t)^{4}>}{2<\Delta {\bm 
\phi}(t)^{2}>^{2}} - 1
\label{rngp}
\end{equation}
where $<\Delta {\bm \phi}(t)^{2n}> = \frac{1}{N}\displaystyle \sum_{i=1}^{N}
<|{\bm \phi}_{i}(t) - {\bm \phi}_{i}(0)|^{2n}>$.

\begin{figure}
\epsfig{file=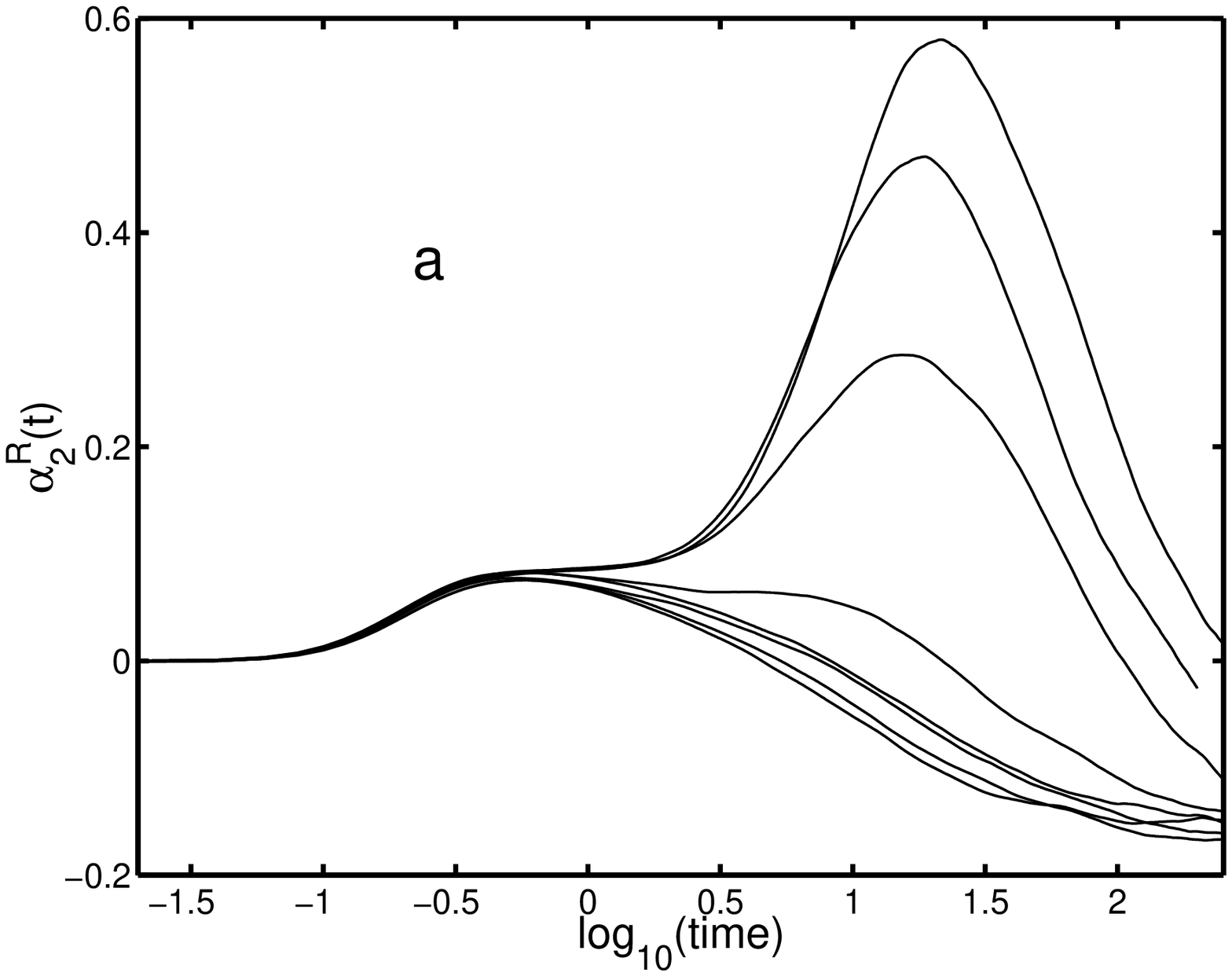,width=7.5cm}
\epsfig{file=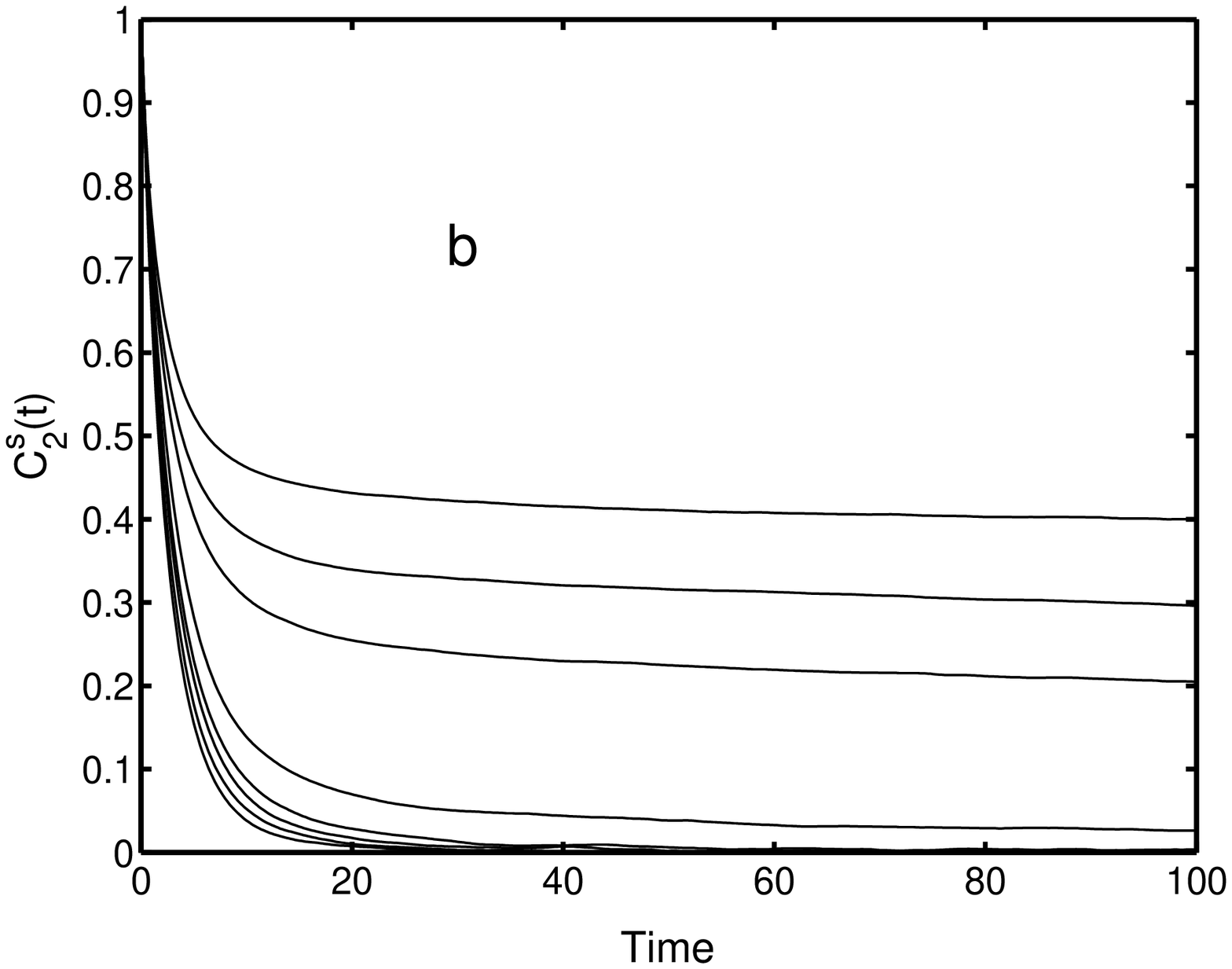,width=7.5cm}
\caption{(a) Semi-log plot of rotational NGP versus the reduced time near 
the I-N transition at different densities ranging from $\rho^* = 0.285$ to 
$\rho^*=0.32$ in a grid of $\delta \rho^*= 0.005$ (arranged in the 
increasing order of density from bottom to top). 
(b) Single particle second rank orientational time correlation function 
versus reduced time for the corresponding densities arranged in the reverse 
order.}
\label{ngtR}
\end{figure}

\begin{figure}
\epsfig{file=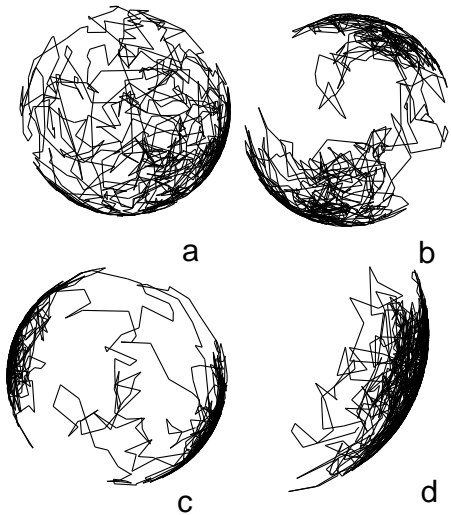,width=6cm}
\caption{Typical single particle trajectories of the unit orientation vector 
at four densities: (a) $\rho^*= 0.285$ (in the isotropic phase), 
(b) $\rho^*=0.31$ (very close to the I-N transition), (c) $\rho^*= 0.315$ 
(at the I-N transition point), and (d) $\rho^*= 0.32$ (in the nematic 
phase).}
\label{ngtR}
\end{figure}

{\bf Figure 1a} shows the time dependence of the rotational NGP 
$\alpha_{2}^{(R)}(t)$ for a series of densities across the I-N transition. 
In this case, the transition density is at $\rho^*_{IN}= 0.315$ when 
temperature $T^*=1$ \cite{Jose-Bagchi}. Note the growth of a dominant peak 
following a shoulder on approaching the I-N transition from the isotropic 
side. {\it The appearance of a shoulder at small times in the rotational NGP
is accompanied by a signature of a sub-diffusive regime in the temporal 
evolution of mean square angular deviation} (data not shown). Also, the time
scale of the shoulder is found to nearly coincide with that of the onset
of the sub-diffusive regime. The shoulder in the rotational NGP can therefore
be ascribed to what may be called the rotational analog of {\it rattling 
within a cage}. Subsequent to the shoulder, the second peak appears around a
time $t^{\star}_{max} \approx 20$, which shifts rather slowly to higher 
values with increasing density. We note that this time is comparable to 
the onset of diffusive motion for rotational degrees of freedom. We have
further explored whether the time scale of the second peak of 
$\alpha_{2}^{(R)}(t)$ has any correlation in the relaxation of the second 
rank single particle orientational time correlation function $C_{2}^{s}(t)$. The latter is defined for a system of ellipsoids as
\begin{equation}
C_{2}^{s}(t)={\frac{\langle{\sum_{i} P_{2}({\bf \hat e}_{i}(0)\cdot 
{\bf \hat e}_{i}(t))}\rangle}{
\langle{\sum_i P_{2}({\bf \hat e}_{i}(0)\cdot {\bf \hat e}_{i}(0))}\rangle}},
\label{cs2}
\end{equation}
where ${\bf \hat e}_{i}$ is the unit vector along the principal axis of 
symmetry of the ellipsoid $i$, $P_{2}$ is the second rank Legendre 
polynomial and the angular brackets stand for statistical averaging.
In {\bf Figure 1b}, the time dependence of $C_{2}^{s}(t)$ is shown for
several densities. We find that the time scale of the second peak also
marks the {\it onset of slow relaxation} in $C_{2}^{s}(t)$. This suggests 
that the second peak is due to the heterogeneity in rotational dynamics
arising from pseudonematic domains that have random local directors. 

In {\bf Figure 2}, we show the microscopic reason for the glassy relaxation
of the nematogens with a display of single particle trajectories in the 
orientational space. This clearly demonstrates the onset of localization of 
the orientational motion in a preferred direction which is the direction of 
the director of a pseudonematic domain. The single particle trajectories 
give the direct evidence of rotational symmetry breaking in the dynamics 
near the I-N transition. Note that the interaction between the ellipsoids 
still has the up-down symmetry; however, the dynamics of the director looses this symmetry in the nematic phase. 

\begin{figure}
\epsfig{file=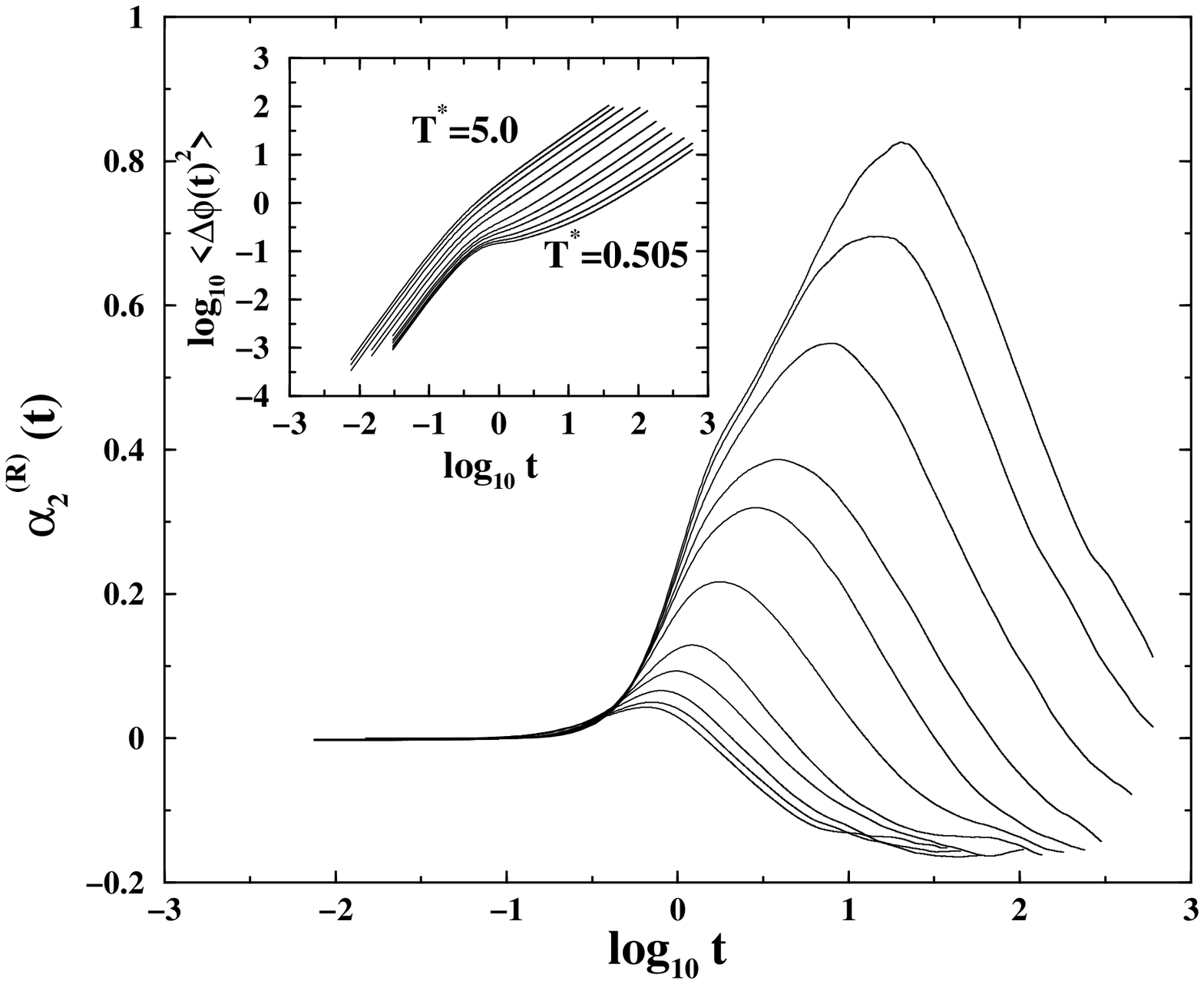,width=7.5cm}
\epsfig{file=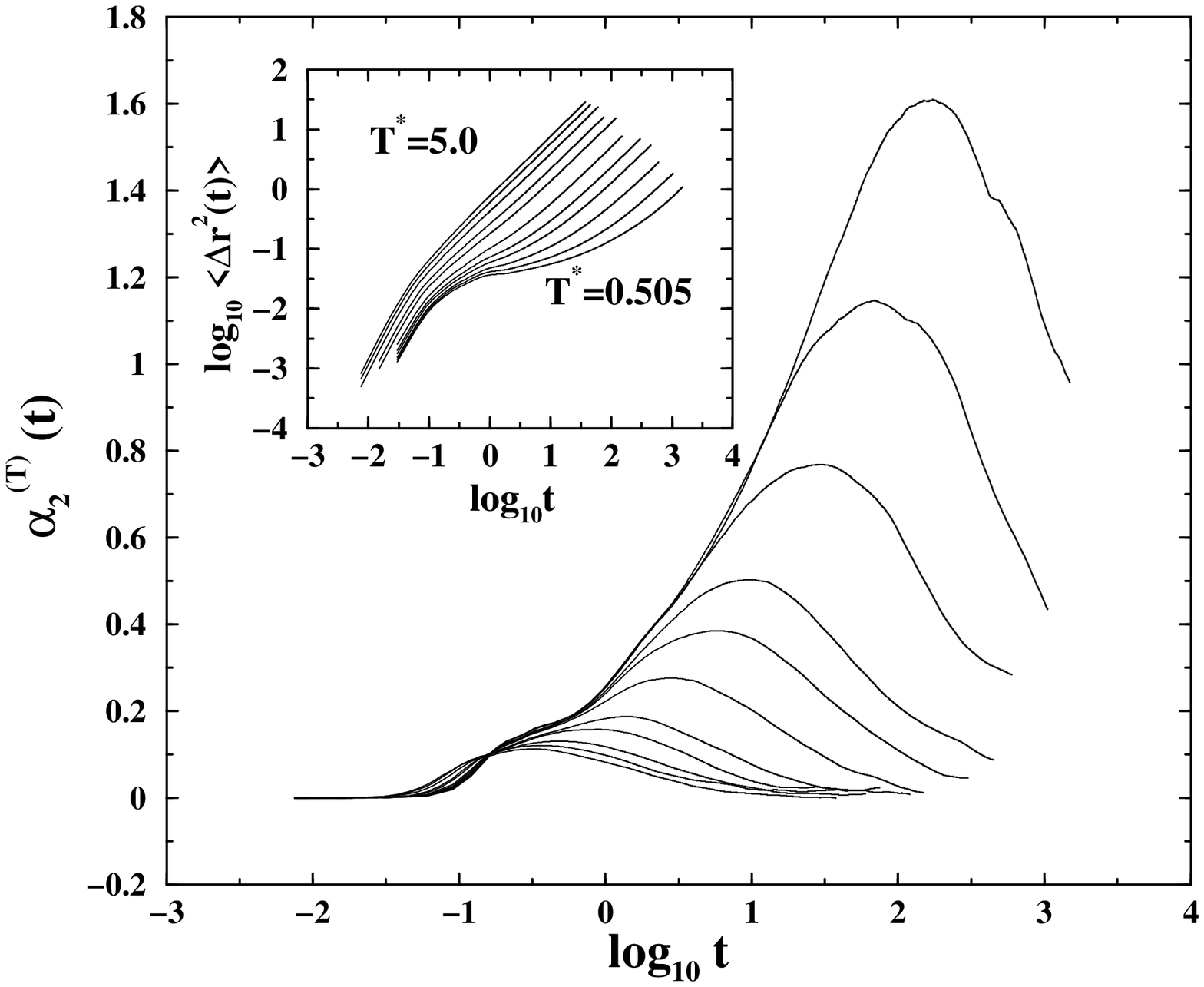,width=7.5cm}
\caption{(a) Semi-log plot of rotational NGP versus the reduced time 
for the ellipsoids in the binary mixture for all temperatures studied. From
bottom to top the temperature decreases from $T^{\star} = 5$ to 
$T^{\star} = 0.505$. The inset shows the time evolution of the mean square
angular deviation (msad) for the same set of temperatures in a log-log plot. 
(b) Semi-log plot for the corresponding translational NGP showing its time 
evolution. The inset displays the time dependence of the mean square 
deviation (msd) in a log-log plot.}
\end{figure}
The above dynamical features have been found to have some correspondence 
with those revealed by the simulations of our binary system in the 
supercooled regime \cite{Chakrbarti-Bagchi}. {\bf Figures 3a} and 
{\bf 3b} show the temporal evolution of the rotational and translational 
NGPs, respectively, of the ellipsoids in the binary mixture for a series of 
temperatures close to the mode-coupling theory critical temperature. 
It is evident from {\bf Figure 3a} that the growth of the rotational NGP 
to its maximum value $\alpha_{2max}^{(R)}$ before it starts decaying is 
rather smooth at all the temperatures studied and the inset shows that the 
maximum is reached on a time scale that characterizes the onset of diffusive
motion for rotational degrees of freedom. This time $t_{max}^{(R)}$ gets 
lengthened with decreasing temperature. The time dependence of the 
translational NGP, however, differs in having a shoulder that resembles that of the rotational 
NGP near the I-N transition. Starting from $\alpha_{2}^{(T)} (t = 0) = 0$, 
translational NGP rises smoothly to a value $\alpha_{2max}^{(T)}$ on a time 
scale that characterizes the crossover from the ballistic to diffusive 
motion for translational degrees of freedom at high temperatures. It then 
starts falling off to reach the long time limit. As temperature drops, 
{\it a shoulder} or {\it a step-like feature} appears in between the initial rise and subsequent growth to its maximum value. We note that the shoulder 
appears as a sub-diffusive regime sets in between the ballistic and 
diffusive motion as evident from the time dependence of the mean square 
displacement shown in the inset of {\bf Figure 3b}. It has been further
observed (data not shown) that the shoulder in translational NGP appears on 
a time scale that coincides with the second maximum of the velocity 
autocorrelation function which shows oscillatory character, particularly at 
low temperatures. The above analysis suggests that the shoulder at small 
times in the translational NGP can be attributed to what is known as the
{\it rattling within a cage} while the dominant peak appears on a time scale that characterizes the escape of a particle from the cage formed by its 
nearest neighbors. The absence of the step-like feature in the time 
dependence of the rotational NGP may be ascribed to the observation that 
the freezing in rotational degrees of freedom is less pronounced than it is 
in translational degrees of freedom \cite{Chakrbarti-Bagchi}. 

That the long time limit of the rotational NGP, defined by Eq. (2), is 
different from its translational counterpart requires further attention. 
Note that there is as yet no unique way of computing NGP 
for rotational degrees of freedom because the angular space is as such 
bounded while one needs the rotational analog of ${\bf r}_{i}(t)$ to be an 
unbounded one. Here we have followed a procedure that has been used 
previously to compute rotational diffusion \cite{rd-PRE}, but there is no 
earlier report of rotational NGP in a three-dimensional system (to the best 
of our knowledge). Although the prefactor of $1/2$ in Eq. (2), which is the 
appropriate one for there being two rotational degrees of freedom for the 
ellipsoids within the linear molecule approximation, can capture the 
expected zero time limit, it does show a deviation from the long time limit 
which one would desire to be zero.

\begin{figure}
\epsfig{file=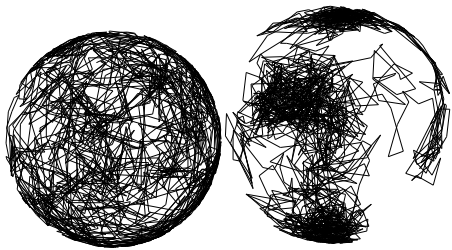,width=6cm}
\caption{Typical trajectories of the orientational unit vector of a single
particle at the highest (left) and lowest (right) temperatures studied.}
\end{figure}
In {\bf Figure 4}, we display single particle trajectories of the ellipsoids
in the binary mixture in the orientational space. While the 
dynamics is ergodic at high temperatures, the signature of non-ergodicity
is evident at the lowest temperature studied. A comparison between Figures
2 and 4 reveals both the similarities and the differences between the 
rotational dynamics of the nematogens (near the I-N transition) and the 
ellipsoids in the binary mixture. For the liquid crystal system 
"nonergodic-like" behavior is driven by the rotational anisotropy that
emerges near the I-N transition. Note the large scale angular displacement
near the I-N transition which is similar to the one observed in the 
supercooled regime \cite{Chakrbarti-Bagchi}.

Mode coupling theory (MCT) may provide valuable insight into the origin of
the slow down in both the cases. MCT predicts that near the I-N transition, 
the low frequency rotational memory kernel should diverge in a power law 
fashion \cite{Gottke-Brace-JCP-2002}. 
\begin{equation}
M_R(z)\approx \frac{A}{z^\alpha}
\end{equation}
with $\alpha \simeq 0.5$. This leads to the power law in the temporal decay 
of the orientational time correlation function at intermediate times. Note, 
however, that the above singularity is not robust in the sense that it 
assumes a divergence of static orientational pair correlation function at 
small wavenumbers ($k\simeq0$), which never occurs because the I-N 
transition intervenes before that \cite{Zwanzig-1963}. For density 
fluctuations in the supercooled liquid, the relevant memory function is that
of the longitudinal current and this function also develops a power-law
singularity of the same type as Eq. (4) \cite{Gotze,Leutheusser}. However, 
the origin of this singularity is due to the formation of pronounced local 
order at molecular length scale (first peak of the static structure factor).
Considering the analogous dynamical behavior revealed in recent studies, an 
attempt to explore the difference in physical origin within the framework of MCT would definitely be a worthwhile undertaking, particularly the 
crossover from the power law to the LdG exponential relaxation near the 
I-N transition.

Perhaps the difference is best understood by looking at the free energy
landscape. While the free energy with respect to orientational
density fluctuations becomes increasingly flat as the 
I-N transition is approached, the free energy surface becomes rugged 
for the supercooled liquid. The confining mean potential for the orientation
is a symmetric double well separated by an angle of $\pi$ at the I-N 
transition. This point, however, deserves further study.

It is a pleasure to thank Dr B. Prabhakar for helpful discussions. This work was supported in parts by grants from DST and CSIR, India. DC acknowledges 
UGC, India for providing Research Fellowship.


\begin{references}
\bibitem{GB-parameter} $\kappa = 3, \kappa^{\prime} = 5, \mu = 2$, and 
$\nu = 1$. All quantities reported here for this system are in reduced 
units, defined in terms of the Gay-Berne potential parameters 
$\epsilon_{GB}$ and $\sigma_{GB}$ and the mass of the ellipsoids, all of
which are set equal to unity. See Ref. \cite{Jose-Bagchi} for details of 
simulation.
\bibitem{binary-parameter} All quantities are given in reduced units 
defined in terms of Lennard-Jones potential parameters $\epsilon_{LJ}$ and
$\sigma_{LJ}$: length in units of $\sigma_{LJ}$, temperature in units of
$\epsilon_{LJ}/k_{B}$, $k_{B}$ being the Boltzmann constant, and time in 
units of $(\sigma_{LJ}^2m/\epsilon_{LJ})^{1/2}$, $m$ being the mass of 
spheres. We set the molecular masses of both spheres and ellipsoids and the 
moment of inertia of the ellipsoids equal to unity. The various energy 
and length parameters of the interaction potentials for our binary 
mixture have been chosen as follows: $\epsilon_{LJ} = 1.0$ , 
$\epsilon_{GB} = 0.5$, $\epsilon_{SE} = 1.5$, $\sigma_{LJ} = 1.0$, 
$\sigma_{GB} = 1.0$, and $\sigma_{SE} = 1.0$. See Ref. 
\cite{Cleaver-PRE-JCP} for the general implication of the symbols. Unlike in
Ref. \cite{Cleaver-PRE-JCP}, here the subscripts are used to distinguish 
between the Lennard-Jones (LJ) interactions, the Gay-Berne (GB) 
interactions, and the sphere-ellipsoid (SE) interactions in the binary 
mixture. For the Gay-Berne interactions, $\kappa = 2, \kappa^{\prime} = 5, 
\mu = 2$, and $\nu = 1$. For the sphere-ellipsoid interaction 
$\epsilon_{E} / \epsilon_{S}  = 0.2$ and $\mu = 2$. With these set of 
parameter values, the binary mixture has been investigated along an isochore
corresponding to the reduced density $\rho^{\star} = 0.8$ at a series of 
reduced temperatures ranging from $5.0$ to $0.505$.
\end{references}
\end{document}